\documentclass[
aps,prl,
amsmath,amssymb,
reprint,%
]{revtex4-1}

\usepackage{graphicx}
\usepackage{dcolumn}
\usepackage{bm}
\usepackage{amsmath}
\usepackage{amssymb}
\usepackage{hyperref}

\newcommand{\be}{\begin{equation}}
\newcommand{\ee}{\end{equation}}
\newcommand{\bea}{\begin{eqnarray}}
\newcommand{\eea}{\end{eqnarray}}

\newcommand{\eqn}[1]{(\ref{#1})}

\newcommand{\cf}{{\cal F}}
\newcommand{\cb}{{}}
\newcommand{\ch}{{\cal H}}
\newcommand{\cn}{{\cal N}}
\newcommand{\cl}{{\cal L}}

\newcommand{\vp}{\varphi}

\begin{document}

\title[Sample title]{Energy loss of a nonaccelerating quark moving through a strongly coupled\\$\mathcal{N}=4$ super Yang-Mills vacuum or plasma in strong magnetic field}

\author{Kiminad A. Mamo}
 \affiliation{Department of Physics, University of Illinois, Chicago, Illinois 60607, USA}


\begin{abstract}
Using AdS/CFT correspondence, we find that a massless quark moving at the speed of light $v=1$, in arbitrary direction, through a strongly coupled $\mathcal{N}=4$ super Yang-Mills (SYM) vacuum at $T=0$, in the presence of strong magnetic field $\mathcal{B}$, loses its energy at a rate linearly dependent on $\mathcal{B}$, i.e., $\frac{dE}{dt}=-\frac{\sqrt{\lambda}}{6\pi}\mathcal{B}$. We also show that a heavy quark of mass $M\neq 0$ moving at near the speed of light $v^2=v_{*}^2=1-\frac{4\pi^2 T^2}{\mathcal{B}}\simeq1$, in arbitrary direction, through a strongly coupled $\mathcal{N}=4$ SYM plasma at finite temperature $T\neq 0$, in the presence of strong magnetic field $\mathcal{B}\gg T^2$, loses its energy at a rate linearly dependent on $\mathcal{B}$, i.e., $\frac{dE}{dt}=-\frac{\sqrt{\lambda}}{6\pi}\mathcal{B}v_{*}^2\simeq-\frac{\sqrt{\lambda}}{6\pi}\mathcal{B}$.  Moreover, we argue that, in the strong magnetic field $\mathcal{B}\gg T^2$ (IR) regime, $\mathcal{N}=4$ SYM and adjoint QCD theories (when the adjoint QCD theory has four flavors of Weyl fermions and is at its conformal IR fixed point $\lambda=\lambda^*$) have the same microscopic degrees of freedom (i.e., gluons and lowest Landau levels of Weyl fermions) even though they have quite different microscopic degrees of freedom in the UV when we consider higher Landau levels. Therefore, in the strong magnetic field $\mathcal{B}\gg T^2$ (IR) regime, the thermodynamic and hydrodynamic properties of $\mathcal{N}=4$ SYM and adjoint QCD plasmas, as well as the rates of energy loss of a quark moving through the plasmas, should be the same.
\end{abstract}
\pacs{11.25.Tq}

\maketitle


\section{I. Introduction} The quark-gluon plasma (QGP) created in ultra-relativistic heavy-ion collisions may be subject to a strong magnetic field produced by many spectator nucleons~\cite{Kharzeev:2007jp}, and recently, the effect of this strong magnetic field on several dynamical~\cite{Bali:2011qj, Mamo:2015dea, Dudal:2015wfn, Fang:2016cnt, Rougemont:2015oea, Li:2016gtz, Evans:2016jzo} and topological~\cite{Fukushima:2008xe,Son:2004tq,Kharzeev:2010gd} properties of QGP, including its thermalization \cite{Fuini:2015hba, Mamo:2015aia}, has been explored. In addition, the effect of the strong magnetic field on the heavy quark and jet quenching parameter has been studied both at weak coupling using QCD \cite{Stephanov:2015roa, Sadofyev:2015tmb,Li:2016bbh} and at strong coupling using $\mathcal{N}$=4 super Yang-Mills (SYM) theory \cite{Li:2016bbh} .

The field content of $\mathcal{N}$=4 SYM theory, including their $U(1)\subset SU(4)$ R-symmetry charge, is as follows (all of them are in adjoint representation of the gauge group $SU(N_{c})$), see for example \cite{D'Hoker:2009mm}: there are four flavors of Weyl fermions (1 Weyl fermion of charge 1 and 3 Weyl fermions of charge $-\frac{1}{3})$; 3 complex scalar field of charge $\frac{2}{3}$; and 1 vector field of charge 0 (the gauge field). And, the spectrum of single particle excitations of $\mathcal{N}$=4 SYM theory in the presence of a magnetic field pointing in the $z$ direction are given by relativistic Landau levels which are the following \cite{D'Hoker:2009mm}: for a charge $q_{\phi}$ scalar field
\begin{equation}\label{scalars}
  E_{n}=\sqrt{\lvert q_{\phi}\mathcal{B}\rvert (2n+1) + p_{z}^2},\,\,\, n=0,1,2,...\,;
\end{equation}
for a charge $q_{\psi}$ Weyl fermion (with $s_{z}=\pm\frac{1}{2}$)
\begin{equation}\label{fermions}
 E_{n}=\sqrt{2\lvert q_{\psi}\mathcal{B}\rvert(n+\frac{1}{2}-s_{z}) + p_{z}^2},\,\,\, n=0,1,2,...\,.
\end{equation}
From \eqn{scalars} and \eqn{fermions} it is clear that in the lowest Landau level (LLL) with zero energy (at vanishing momentum $p_{z}$) we only have Weyl fermions but no scalars. Hence, in the strong magnetic field $\mathcal{B}\gg T^2$ regime the whole dynamics of $\mathcal{N}=4$ SYM theory is entirely dominated by the lowest Landau levels (LLLs) of Weyl fermions with four flavors (in the adjoint representation) since the scalar particles (and higher Landau levels of Weyl fermions) are integrated out in this regime resulting in a (1+1)-dimensional low energy effective field theory of LLLs and the gauge field.

In contrast, the field content of adjoint QCD with four flavors, including their $U(1)\subset SU(4)$ flavor-symmetry charge, is as follows (all of them are in adjoint representation of the gauge group $SU(N_{c})$), see for example \cite{Shifman:2013yca}: there are four flavors of Weyl fermions (1 Weyl fermion of charge 1 and 3 Weyl fermions of charge $-\frac{1}{3})$; and 1 vector field of charge 0 (the gauge field). And, the spectrum of single particle excitations of adjoint QCD in the presence of a magnetic field pointing in the $z$ direction are given by the relativistic Landau levels which
for a charge $q_{\psi}$ Weyl fermion (with $s_{z}=\pm\frac{1}{2}$) are given by
\begin{equation}\label{QCDfermions}
 E_{n}=\sqrt{2\lvert q_{\psi}\mathcal{B}\rvert(n+\frac{1}{2}-s_{z}) + p_{z}^2},\,\,\, n=0,1,2,...\,.
\end{equation}
Hence, in the strong magnetic field $\mathcal{B}\gg T^2$ regime the whole dynamics of adjoint QCD is entirely dominated by the lowest Landau levels (LLLs) of Weyl fermions with four flavors (in the adjoint representation) since the higher Landau levels of Weyl fermions are integrated out in this regime resulting in a (1+1)-dimensional low energy effective field theory of LLLs and the gauge field. Note that the beta function for adjoint QCD with four flavors is given by \cite{Shifman:2013yca}, see also \cite{Jones,Caswell},
\begin{equation}\label{beta}
  \beta=\mu\frac{\partial}{\partial\mu}\lambda(\mu)\equiv-\frac{1}{2}\frac{\lambda^2}{(2\pi)^2}+\frac{5}{4}\frac{\lambda^3}{(2\pi)^4}\,,
\end{equation}
which has vanishing beta function or IR fixed point at $g^2_{YM}N_{c}\equiv\lambda=\lambda^*=\frac{8}{5}\pi^2$. Since, the beta function of $\mathcal{N}$=4 SYM vanishes for any 't Hooft coupling $\lambda$, we can claim that

\begin{center}
  \emph{$\mathcal{N}=4$ SYM in strong magnetic field $\mathcal{B}\gg T^2$ at $\lambda=\lambda^*$ $\equiv$ adjoint QCD with four flavors in strong magnetic field $\mathcal{B}\gg T^2$} at $\lambda=\lambda^*$,
\end{center}
where $\lambda^*$ is defined as the coupling at which the beta function of adjoint QCD with four flavors vanishes. Note that in this article whenever we refer to adjoint QCD we are specifically referring to the adjoint QCD with four flavors and at its conformal IR fixed point $\lambda=\lambda^*$.

Due to the above equivalence, using the AdS/CFT correspondence in order to study the effect of the strong magnetic field $\mathcal{B}\gg T^2$ on a strongly coupled $\mathcal{N}$=4 SYM plasma or vacuum is particularly interesting, since the results found for $\mathcal{N}=4$ SYM (at strong coupling and large $N_{c}$ limit) also apply for adjoint QCD (at strong coupling and large $N_{c}$ limit). Therefore, we can conclude that the entropy density \cite{D'Hoker:2009mm}
\begin{equation}\label{s}
 s=\frac{1}{3\sqrt{3}}N_{c}^2\mathcal{B}T\,,
\end{equation}
conductivity \cite{Mamo:2013efa}
\begin{equation}\label{sigma}
 \sigma^{\parallel}=\frac{1}{32\sqrt{3}\pi^3}\frac{\mathcal{B}}{T}\,,
\end{equation}
shear viscosity to entropy density ratio \cite{Critelli:2014kra}
\begin{equation}\label{eta}
  \frac{\eta^{\parallel}}{s}=\pi\frac{T^2}{\mathcal{B}}\,,
\end{equation}
and Chern-Simons diffusion rate \cite{Basar:2012gh}
\begin{equation}\label{CS}
\Gamma=\frac{\lambda^2}{384\sqrt{3}\pi^5}\mathcal{B}T^2\,,
\end{equation}
of $\mathcal{N}=4$ SYM plasma, in the strong magnetic field $\mathcal{B}\gg T^2$ regime, are also the entropy density, conductivity, shear viscosity to entropy density ratio, and Chern-Simons diffusion rate of adjoint QCD plasma in strong magnetic field
$\mathcal{B}\gg T^2$ at $\lambda=\lambda^*$.

It would be very interesting to check the above claim numerically using the lattice adjoint QCD \cite{DeGrand:2011qd} in strong magnetic field $\mathcal{B}\gg T^2$ regime (for lattice QCD in magnetic field see \cite{Bali:2014kia, Bali:2011qj}) which would also be a nice numerical verification of the AdS/CFT correspondence in a set up where supersymmetry is totally broken unlike the previous numerical tests of the AdS/CFT correspondence which rely on supersymmetry \cite{Kadoh:2016eju}.

\section{II. Energy and momentum loss} It is well known that a quark moving at a constant velocity $v$, for example, through a strongly coupled $\mathcal{N}$=4 SYM vacuum, doesn't loss its energy, even though it does in a plasma at finite temperature $T$. The rates of energy and momentum loss of a heavy quark moving at constant velocity $v$ through a strongly coupled $\mathcal{N}$=4 SYM plasma, with no magnetic field, were first computed in \cite{Gubser:2006bz, Herzog:2006gh} using the AdS/CFT correspondence \cite{Maldacena:1997re, Gubser:1998bc, Witten:1998qj}. Effects of fluid velocity gradients and axial chemical potential on heavy quark energy loss has also been investigated in \cite{Rajagopal:2015roa, Lekaveckas:2013lha}. And, the rates of energy and momentum loss of an accelerating quark moving through a strongly coupled $\mathcal{N}$=4 SYM vacuum, with no magnetic field, was found in \cite{Mikhailov:2003er, Chernicoff:2008sa, Fadafan:2008bq}, see also \cite{Fiol:2012sg, Lewkowycz:2013laa, Banerjee:2013rca, Banerjee:2015fed}.

In this article, using the AdS/CFT correspondence, we show that in the presence of a strong magnetic field $\mathcal{B}$, even a nonaccelerating quark moving at a constant velocity $v$, through a strongly coupled $\mathcal{N}$=4 SYM vacuum at $T=0$, loses its energy at a rate linearly dependent on $\mathcal{B}$.

We will study the rates of energy and momentum loss of a heavy quark of mass $M$ moving with velocity $v$, in arbitrary direction, through a strongly magnetized plasma in the strong coupling regime. The effect of the magnetic field directly on the heavy quark moving through a non-magnetized plasma (ignoring the effect of the magnetic field on the plasma) was studied in \cite{Matsuo:2006ws, Kiritsis:2011ha}. In this article, we rather ignore the effect of the magnetic field $\mathcal{B}$ directly on the heavy quark of relativistic mass $\gamma M\gg \sqrt{\mathcal{B}}$, where the Lorentz factor $\gamma=\frac{1}{\sqrt{1-v^2}}$, and only consider the effect of the strong magnetic field $\mathcal{B}\gg T^2$ on the plasma. In other words, we will work on the more physical limit $\gamma M\gg \sqrt{\mathcal{B}}\gg T$.

Specifically, we will study the rates of energy and momentum loss of a heavy quark of mass $M$ moving at constant velocity $v$ through a strongly coupled $\cn=4$ SYM plasma in the presence of strong magnetic field $\mathcal{B}\gg T^2$ using its 5-dimensional gravity dual.

The 5-dimensional background metric in the presence of strong magnetic field $\mathcal{B}\gg T^2$ is given by \cite{D'Hoker:2009mm},
\bea \label{BTZu}
ds^{2}&=&g_{MN}dx^{M}dx^{N}=\frac{\mathcal{R}^2}{u^2}\left(-f(u)dt^2+dz^2\right)\nonumber\\
&+&\mathcal{R}^2\mathcal{B}(dx^2+dy^2)+\frac{\mathcal{R}^2}{u^2f(u)}du^2\,,
\eea
where $f(u)=1-\frac{u^2}{u_{h}^2}$, the horizon corresponds to $u=u_{h}$, the boundary to $u=0$, the Hawking temperature $T$ of the BTZ black hole is $T=\frac{1}{2\pi u_{h}}$, we identify $\mathcal{R}=\frac{R}{\sqrt{3}}$ as the radius of the $AdS_{3}$ spacetime or BTZ black hole, and $\mathcal{B}=\sqrt{3}eB=\sqrt{3}F_{xy}$ as the physical magnetic field at the boundary. Also, note that \eqn{BTZu} is valid only near the horizon, i.e., in the regime $u\gg u_{0}=\frac{1}{\sqrt{\mathcal{B}}}$. And, for an arbitrary strength of $\mathcal{B}$, the metric numerically interpolates between the $AdS_{3}$ spacetime or BTZ black hole \eqn{BTZu} near the horizon (IR) and $AdS_{5}$ spacetime near the boundary (UV)\cite{D'Hoker:2009mm}.

We will further rewrite the metric \eqn{BTZu} as
\be \label{BTZm}
ds^2=\frac{\mathcal{R}^2}{u^2}\Big( -\cf\, dt^2+dz^2
+ \ch (dx^2+dy^2) +\frac{ du^2}{\cf}\Big),
\ee
where $\cf=f(u)=1-\frac{u^2}{u_{h}^2}$, and $\ch=u^2\mathcal{B}$, so that, it resembles the anisotropic metric used in \cite{Chernicoff:2012iq}, which we will follow closely in the following derivation of the energy and momentum loses of a heavy quark.

On the gravity side the rates of energy and momentum loss of a heavy quark are described by a string propagating in the background \eqn{BTZm} governed by the equation of motion for the string which is derived from the Nambu-Goto action
\be
S =- \frac{1}{2\pi\alpha'}\int d\tau d\sigma\,  \sqrt{-det\,h_{ab}} =
\int d\tau d\sigma \, {\cal L}  \,,
\label{stringaction}
\ee
where $h_{ab}=g_{MN}\partial_{a}X^{M}(\tau,\sigma)\partial_{b}X^{N}(\tau,\sigma)$ is the induced worldsheet metric. In the following expressions, we set $\mathcal{R}^2/2\pi\alpha'=\sqrt{\lambda}/6\pi$ to one, and reinstate it at the end.

From the action \eqn{stringaction}, we determine the spacetime momentum flow $\Pi_M$ along the string to be
\be
\Pi_M = \frac{\partial \cl}{\partial (\partial_\sigma X^M)} \,.
\ee

Since, we have rotational symmetry in the $xy$-directions, we can set $y=0$. Then, identifying $(t,u)=(\tau,\sigma)$ and considering a string embedding of the form
\bea
x(t,u)&\to&  \Big( vt + x(u) \Big) \cos \varphi \,,\\
z(t,u)&\to&  \Big( vt + z(u) \Big) \sin \varphi \,,
\label{pro}
\eea
which corresponds to a quark moving with velocity $v$ in the $xz$-plane at an angle $\varphi$ with the $x$-axis, the Lagrangian takes the form
\bea
\cl &=& - \frac{1}{u^2\sqrt{\cf}}\Bigg[\cb \cf +  \sin^2 \varphi \, ( \cb \cf^2 z'^2 -v^2)\nonumber\\
&+&
\ch \cos^2 \varphi \, \Big[ \cb \cf^2 x'^2 -v^2 - \cf v^2 (z' - x')^2 \sin^2
\varphi \Big]  \Bigg]^{1/2} \,,
\eea
and, the rates at which energy and momentum flow from the boundary to the horizon along the string become
\bea
- \Pi_t &=& \frac{1}{\cl u^{4}} \,\cb \cf v \Big[  x' \sin^2 \vp + \ch z' \cos^2 \vp \Big]
\,,\nonumber \\ [1.3mm]
\Pi_x &=& \frac{1}{\cl u^{4}} \, \ch \Big[  \cb \cf\,  x' + v^2 (z' - x') \sin^2 \varphi \Big] \cos \varphi  \,,\nonumber \\ [1.3mm]
\Pi_z &=& \frac{1}{\cl u^{4}} \,\Big[ \cb \cf \,z' + \ch v^2 (x' -z') \cos^2 \varphi \Big] \sin \varphi  \,,
\label{momenta}
\eea
where $'$ denotes differentiation with respect to $u$. Note that
\be
- \Pi_t = \Pi_x \, v \cos \vp + \Pi_z \, v \sin \vp = \vec{\Pi}\cdot \vec{v}\,.
\ee

Moreover, from the equation of motion $\partial_{u}\Pi_M=0$ (which is valid only when the end of the string or the heavy quark is nonaccelerating), we find that $\Pi_M$ is a constant independent of $u$ or the mass of the quark $M=\frac{\sqrt{\lambda}}{2\pi}\big(\frac{1}{u}-\frac{1}{u_{c}}\big)$ \cite{Herzog:2006gh} where $u$ is the radial location at which the end of the string is attached to, say, a D7 brane. And, $u_{c}$ is the radius of the worldsheet horizon with $u_{c}=u_{h}$ for $v=0$, and constrained by $u\leq u_{c}$ which is determined by requiring the time-time component of the worldsheet metric to always be negative or zero. Therefore, we are free to fix $M$ to any value as long as it satisfies the bound $\sqrt{\lambda}\sqrt{\mathcal{B}}\gg M\gg \sqrt{\lambda}T$ which is the result of the geometrical bound $u_{0}\ll u\ll u_{h}$, on the gravity side, and the physical requirement that the mass of the heavy quark $M$ must be much larger than the temperature $T$ of the plasma, i.e., $M\gg \sqrt{\lambda}T$ or $u\ll u_{h}$, so that the heavy quark can be considered a legitimate external probe of the plasma.

In addition, we should note that, since requiring the time-time component of the worldsheet metric at $T=0$ and $B=0$ (for the pure $AdS_5$ bulk metric) to always be negative or zero would result in the constraint $1-v^2\geq0$, we could conclude that the $u=u_c$ or $M=0$ limit must be accompanied by the $v=1$ limit. So, in the vacuum at $T=0$, the bound on $M$ becomes $\sqrt{\lambda}\sqrt{\mathcal{B}}\gg M\geq 0$, hence we are free to set the mass of the quark $M=0$, if we would like to, as long as we also set its velocity $v=1$.


In order to find the background solution of the string, we invert the relations \eqn{momenta} to find
\be
x' = \pm \frac{v}{\cf \sqrt{\cb \ch}} \frac{N_x}{\sqrt{N_z N_x-D}}\,, \,\,\,\,
z' =\pm \frac{\ch v}{\cf \sqrt{\cb \ch}} \frac{N_z}{\sqrt{N_z N_x-D}} \,,
\label{primes}
\ee
where
\bea
N_x&=& -\Pi_x (\cb \cf  \sec \vp - \ch v^2 \cos \vp) + \Pi_z \, \ch v^2 \sin \vp
\,, \\[1.7mm]
N_z&=& -\Pi_z (\cb \cf  \csc \vp -  v^2 \sin \vp)+\Pi_x v^2 \cos \vp \,, \\[1.7mm]
D&=& \frac{\cb \cf  \csc \vp \sec \vp}{u^4}
\Big[  \Pi_z \Pi_x u^4 - \ch v^2 \cos \vp \sin \vp \Big]\nonumber\\
&\times&\Big[ \cb \cf - v^2 \Big( \ch \cos^2 \vp + \sin^2 \vp \Big) \Big] \,.\,\,\,\,\,\,\,\,\,\,
\label{nnd}
\eea
Since, $\cb \cf$ ($\ch$) is monotonically decreasing (increasing) from the boundary to the horizon, the last factor in square brackets in \eqn{nnd} is positive at the boundary and negative at the horizon. Therefore, there exists a critical value $u_c$ in between such that
\be
 \cf_c - v^2 \left( \ch_c \cos^2 \vp + \sin^2 \vp \right) = 0\,,
 \label{hc}
 \ee
where $\ch_c=\ch(u_c)$, and $\cf_c=\cf(u_{c})$. Note that at $u=u_{c}$, $D=0$, and
\be
\left. N_z N_x \right|_{u_c} =
- v^4 \left( \ch_c\,  \Pi_z \cos \vp - \Pi_x \sin \vp \right)^2 \,,
\ee
is negative unless the momenta are related through
\be
\frac{\Pi_z}{\Pi_x} = \frac{\tan \vp}{\ch_c}\,,
\label{relation}
\ee
in which case it vanishes.

Then, requiring the first square bracket in \eqn{nnd} also vanishes at $u=u_c$, and using \eqn{relation}, we find
\be
\Pi_x =  \ch_c\,  \frac{v  \cos \vp}{u_c^2}\,,\,\,\,\,\,\,\,\,\,\,\,
\Pi_z =  \frac{v \sin \vp}{u_c^2} \,.
\label{result}
\ee

Therefore, the drag force or the rate of momentum loss of a heavy quark, defined as $\vec F_{drag}=\frac{d\vec{p}}{dt} \equiv (-\Pi_x, -\Pi_z) $ is (after reinstating the factor $\mathcal{R}^2/2\pi\alpha'=\sqrt{\lambda}/6\pi$)
\be
\vec F_{drag} = -\frac{\sqrt{\lambda}}{6\pi}\,
\frac{v}{u_c^2} \, (\ch_c \cos \vp, \sin \vp) \,,
\label{drag1}
\ee
which is exactly Eq. 3.22 in \cite{Chernicoff:2012iq}, up to an overall minus sign, once we exchange the $x$ and $z$ components of the drag force.

Solving \eqn{hc} for $u_{c}$, we find
\be \label{uc}
u_{c}^2=\frac{1}{\mathcal{B}}\Bigg(\frac{1-v^2\sin^2\vp}{\frac{4\pi^2T^2}{\mathcal{B}}+v^2\cos^2\vp}\Bigg),
\ee
which can be used in \eqn{drag1}, to find
\be \label{drag2}
\vec F_{drag} = -\frac{\sqrt{\lambda}\mathcal{B}v}{6\pi}\Bigg(\cos \vp, \sin \vp\Bigg(\frac{\frac{4\pi^2T^2}{\mathcal{B}}+v^2\cos^2 \vp}{1-v^2\sin^2\vp}\Bigg)\Bigg) \,.
\ee
Note that \eqn{drag2}, exactly reduces to Eq. 3.113 and Eq. 3.95 of \cite{Li:2016bbh}, by the current author, Li, and Yee, when $\vp=0$ (which corresponds to a heavy quark moving in the $x$-direction or perpendicular to the magnetic field) and $\vp=\pi/2$ (which corresponds to a heavy quark moving in the $z$-direction or parallel to the magnetic field), respectively.

In the vacuum at $T=0$, the rate of momentum loss $\frac{d\vec{p}}{dt}$ \eqn{drag2} reduces to
\be \label{drag3}
\frac{d\vec{p}}{dt} = -\frac{\sqrt{\lambda}\mathcal{B}v}{6\pi}\Bigg(\cos \vp, \sin \vp\Bigg(\frac{v^2\cos^2 \vp}{1-v^2\sin^2\vp}\Bigg)\Bigg) \,,
\ee
which for $v=1$ (and $M=0$) becomes

\be \label{drag4}
\frac{d\vec{p}}{dt} = -\frac{\sqrt{\lambda}\mathcal{B}}{6\pi}\big(\cos \vp, \sin \vp\big)\,.
\ee
Therefore, the rate of energy loss $\frac{dE}{dt}=\Pi_t=\frac{d\vec{p}}{dt}\cdot \vec{v}$, for a massless quark moving at the speed of light $v=1$ in $\mathcal{N}$=4 SYM vacuum at $T=0$, is
\be \label{eloss1}
\frac{dE}{dt}=-\frac{\sqrt{\lambda}\mathcal{B}}{6\pi}\,.
\ee

Similarly, for $T\neq 0$ but $\mathcal{B}\gg T^2$ and $v^2=v_{*}^2=1-\frac{4\pi^2T^2}{\mathcal{B}}$, the drag force \eqn{drag2} reduces to
\be \label{drag4}
\vec F_{drag} = \frac{d\vec{p}}{dt} = -\frac{\sqrt{\lambda}\mathcal{B}v_{*}}{6\pi}\big(\cos \vp, \sin \vp\big)=-\frac{\sqrt{\lambda}\mathcal{B}}{6\pi}\vec v_{*}\,.
\ee
Therefore, the rate of energy loss $\frac{dE}{dt}=\Pi_t=\vec F_{drag}\cdot \vec{v}$, for a heavy quark of mass $M$ moving at near the speed of light $v^2=v_{*}^2=1-\frac{4\pi^2T^2}{\mathcal{B}}\simeq 1$ in $\mathcal{N}$=4 SYM plasma at $T\ll \sqrt{\mathcal{B}}$, becomes
\be \label{eloss2}
\frac{dE}{dt}=-\frac{\sqrt{\lambda}\mathcal{B}}{6\pi}v_{*}^2\simeq-\frac{\sqrt{\lambda}\mathcal{B}}{6\pi}\,.
\ee

\section{III. Summary} We have found that a massless quark moving at the speed of light $v=1$, in arbitrary direction, through a strongly coupled and magnetized $\mathcal{N}$=4 SYM vacuum at $T=0$ loses its energy at a rate linearly dependent on $\mathcal{B}$ \eqn{eloss1}
\be \label{eloss3}
\frac{dE}{dt}=-\frac{\sqrt{\lambda}}{6\pi}\mathcal{B}\,.
\ee
We have also found that a heavy quark moving at near the speed of light $v\simeq 1$, in arbitrary direction, through a strongly coupled and magnetized $\mathcal{N}$=4 SYM plasma at $T\neq 0$ loses its energy at a rate linearly dependent on $\mathcal{B}$ \eqn{eloss2}
\be \label{eloss4}
\frac{dE}{dt}\simeq-\frac{\sqrt{\lambda}}{6\pi}\mathcal{B}\,.
\ee

We should also note that the results found in this article for $\mathcal{N}=4$ SYM \eqn{eloss3} and \eqn{eloss4} are also the results one would find for adjoint QCD with four flavors and at IR fixed point $\lambda=\lambda^*$.

From the phenomenological point of view the results found in this article \eqn{eloss3} and \eqn{eloss4} are also very interesting since knowing the rate of energy loss in the presence of a strong magnetic field $B$ is crucial for a complete understanding and numerical simulations of the energy loss mechanisms of the hard probes of the quark-gluon plasma (QGP) produced in heavy ion collisions.

\section{Acknowledgments} The author would like to thank Andrey V. Sadofyev, and Ho-Ung Yee for
helpful discussions and helpful comments on the draft; and Dimitrios Giataganas for comments and enlightening questions on the preprint. The author also would like to thank Tigran Kalaydzhyan, and Mikhail A. Stephanov for discussions. This material is based upon work partially supported by the U.S. Department of Energy, Office of Science, Office of Nuclear Physics, within the framework of the Beam Energy Scan Theory (BEST) Topical Collaboration.


\bibliographystyle{prsty}
\bibliography{ar,tft,qft,books}

\begin{thebibliography}{10}
\bibitem{Kharzeev:2007jp}
  D.~E.~Kharzeev, L.~D.~McLerran and H.~J.~Warringa,
  ``The Effects of topological charge change in heavy ion collisions: 'Event by event P and CP violation',''
  Nucl.\ Phys.\ A {\bf 803}, 227 (2008)
  [arXiv:0711.0950 [hep-ph]].


  \bibitem{Bali:2011qj}
  G.~S.~Bali, {\it et al},
  ``The QCD phase diagram for external magnetic fields,''
  JHEP {\bf 1202}, 044 (2012)
  [arXiv:1111.4956 [hep-ph]].


\bibitem{Mamo:2015dea}
  K.~A.~Mamo,
  ``Inverse magnetic catalysis in holographic models of QCD,''
  JHEP {\bf 1505}, 121 (2015)
 [arXiv:1501.03262 [hep-ph]].

\bibitem{Dudal:2015wfn}
  D.~Dudal, D.~R.~Granado and T.~G.~Mertens,
  ``No inverse magnetic catalysis in the QCD hard and soft wall models,''
  arXiv:1511.04042 [hep-th].

\bibitem{Fang:2016cnt}
  Z.~Fang,
  ``Anomalous dimension, chiral phase transition and inverse magnetic catalysis in soft-wall AdS/QCD,''
  Phys.\ Lett.\ B {\bf 758}, 1 (2016).

\bibitem{Rougemont:2015oea}
  R.~Rougemont, R.~Critelli and J.~Noronha,
  ``Holographic calculation of the QCD crossover temperature in a magnetic field,''
  Phys.\ Rev.\ D {\bf 93}, no. 4, 045013 (2016)
  doi:10.1103/PhysRevD.93.045013
  [arXiv:1505.07894 [hep-th]].

\bibitem{Li:2016gtz}
  S.~w.~Li and T.~Jia,
  ``Dynamically flavored description of holographic QCD in the presence of a magnetic field,''
  arXiv:1604.07197 [hep-th].

\bibitem{Evans:2016jzo}
  N.~Evans, C.~Miller and M.~Scott,
  ``Inverse Magnetic Catalysis in Bottom-Up Holographic QCD,''
  arXiv:1604.06307 [hep-ph].


\bibitem{Fukushima:2008xe}
K.~Fukushima, D.~E. Kharzeev, and H.~J. Warringa,
{The Chiral Magnetic Effect},
  Phys.\ Rev.\ D {\bf D78}, 074033 (2008)
 [arXiv:0808.3382 [hep-ph]].

\bibitem{Son:2004tq}
  D.~T.~Son and A.~R.~Zhitnitsky,
  ``Quantum anomalies in dense matter,''
  Phys.\ Rev.\ D {\bf 70}, 074018 (2004)
  [hep-ph/0405216].

\bibitem{Kharzeev:2010gd}
  D.~E.~Kharzeev and H.~U.~Yee,
  ``Chiral Magnetic Wave,''
  Phys.\ Rev.\ D {\bf 83}, 085007 (2011)
  [arXiv:1012.6026 [hep-th]].

\bibitem{Fuini:2015hba}
  J.~F.~Fuini and L.~G.~Yaffe,
  ``Far-from-equilibrium dynamics of a strongly coupled non-Abelian plasma with non-zero charge density or external magnetic field,''
  JHEP {\bf 1507}, 116 (2015)
  [arXiv:1503.07148 [hep-th]].

\bibitem{Mamo:2015aia}
  K.~A.~Mamo and H.~U.~Yee,
  ``Thermalization of Quark-Gluon Plasma in Magnetic Field at Strong Coupling,''
  Phys.\ Rev.\ D {\bf 92}, no. 10, 105005 (2015)
  [arXiv:1505.01183 [hep-ph]].

\bibitem{Stephanov:2015roa}
  M.~A.~Stephanov and H.~U.~Yee,
  ``No-Drag Frame for Anomalous Chiral Fluid,''
  Phys.\ Rev.\ Lett.\  {\bf 116}, no. 12, 122302 (2016)
  doi:10.1103/PhysRevLett.116.122302
  [arXiv:1508.02396 [hep-th]].

\bibitem{Sadofyev:2015tmb}
  A.~V.~Sadofyev and Y.~Yin,
  ``Chiral Magnetic "Superfluidity",''
  [arXiv:1511.08794 [hep-th]].

\bibitem{Li:2016bbh}
  S.~Li, K.~A.~Mamo and H.~U.~Yee,
  ``Jet quenching parameter of quark-gluon plasma in strong magnetic field: perturbative QCD and AdS/CFT correspondence,''
  [arXiv:1605.00188 [hep-ph]].

\bibitem{D'Hoker:2009mm}
  E.~D'Hoker and P.~Kraus,
  ``Magnetic Brane Solutions in AdS,''
  JHEP {\bf 0910}, 088 (2009)
  [arXiv:0908.3875 [hep-th]].

\bibitem{Shifman:2013yca}
  M.~Shifman,
  ``Remarks on Adjoint QCD with $k$ Flavors, $k\geq 2$,''
  Mod.\ Phys.\ Lett.\ A {\bf 28}, 1350179 (2013)
  [arXiv:1307.5826 [hep-th]].

\bibitem{Jones}
D. R. T. Jones,
``Two-loop diagrams in Yang-Mills theory,''
Nucl. Phys. B 75, 531 (1974).

\bibitem{Caswell}
W. E. Caswell,
``Two-loop diagrams in Yang-Mills theory,''
Phys.Rev. Lett. 33, 244 (1974).


\bibitem{Mamo:2013efa}
  K.~A.~Mamo,
  ``Enhanced thermal photon and dilepton production in strongly coupled $N$ = 4 SYM plasma in strong magnetic field,''
  JHEP {\bf 1308}, 083 (2013)
  [arXiv:1210.7428 [hep-th]].

\bibitem{Critelli:2014kra}
  R.~Critelli, S.~I.~Finazzo, M.~Zaniboni and J.~Noronha,
  ``Anisotropic shear viscosity of a strongly coupled non-Abelian plasma from magnetic branes,''
  Phys.\ Rev.\ D {\bf 90}, no. 6, 066006 (2014)
  [arXiv:1406.6019 [hep-th]].

\bibitem{Basar:2012gh}
  G.~Basar and D.~E.~Kharzeev,
  ``The Chern-Simons diffusion rate in strongly coupled N=4 SYM plasma in an external magnetic field,''
  Phys.\ Rev.\ D {\bf 85}, 086012 (2012)
  [arXiv:1202.2161 [hep-th]].

\bibitem{DeGrand:2011qd}
  T.~DeGrand, Y.~Shamir and B.~Svetitsky,
  ``Infrared fixed point in SU(2) gauge theory with adjoint fermions,''
  Phys.\ Rev.\ D {\bf 83}, 074507 (2011)
  [arXiv:1102.2843 [hep-lat]].

\bibitem{Bali:2014kia}
  G.~S.~Bali, F.~Bruckmann, G.~Endrödi, S.~D.~Katz and A.~Schäfer,
  ``The QCD equation of state in background magnetic fields,''
  JHEP {\bf 1408}, 177 (2014)
  [arXiv:1406.0269 [hep-lat]].

\bibitem{Kadoh:2016eju}
  D.~Kadoh,
  ``Recent progress in lattice supersymmetry: from lattice gauge theory to black holes,''
  PoS LATTICE {\bf 2015}, 017 (2015)
  [arXiv:1607.01170 [hep-lat]].

\bibitem{Gubser:2006bz}
  S.~S.~Gubser,
  ``Drag force in AdS/CFT,''
  Phys.\ Rev.\ D {\bf 74}, 126005 (2006)
  [hep-th/0605182].

\bibitem{Herzog:2006gh}
  C.~P.~Herzog, A.~Karch, P.~Kovtun, C.~Kozcaz and L.~G.~Yaffe,
  ``Energy loss of a heavy quark moving through N=4 supersymmetric Yang-Mills plasma,''
  JHEP {\bf 0607}, 013 (2006)
  [hep-th/0605158].

\bibitem{Maldacena:1997re}
  J.~M.~Maldacena,
  ``The Large N limit of superconformal field theories and supergravity,''
  Adv.\ Theor.\ Math.\ Phys.\  {\bf 2}, 231 (1998)
  [hep-th/9711200].

\bibitem{Gubser:1998bc}
  S.~S.~Gubser, I.~R.~Klebanov and A.~M.~Polyakov,
  ``Gauge theory correlators from noncritical string theory,''
  Phys.\ Lett.\ B {\bf 428}, 105 (1998)
  [hep-th/9802109].

\bibitem{Witten:1998qj}
  E.~Witten,
  ``Anti-de Sitter space and holography,''
  Adv.\ Theor.\ Math.\ Phys.\  {\bf 2}, 253 (1998)
  [hep-th/9802150].

\bibitem{Rajagopal:2015roa}
  K.~Rajagopal and A.~V.~Sadofyev,
  ``Chiral drag force,''
  JHEP {\bf 1510}, 018 (2015)
  [arXiv:1505.07379 [hep-th]].
  
\bibitem{Lekaveckas:2013lha}
M.~Lekaveckas and K.~Rajagopal,
``Effects of Fluid Velocity Gradients on Heavy Quark Energy Loss,''
JHEP {\bf 1402}, 068 (2014)
[arXiv:1311.5577 [hep-th]].

\bibitem{Mikhailov:2003er}
  A.~Mikhailov,
  ``Nonlinear waves in AdS / CFT correspondence,''
  hep-th/0305196.

\bibitem{Chernicoff:2008sa}
  M.~Chernicoff and A.~Guijosa,
  ``Acceleration, Energy Loss and Screening in Strongly-Coupled Gauge Theories,''
  JHEP {\bf 0806}, 005 (2008)
  [arXiv:0803.3070 [hep-th]].

\bibitem{Fadafan:2008bq}
  K.~B.~Fadafan, H.~Liu, K.~Rajagopal and U.~A.~Wiedemann,
  ``Stirring Strongly Coupled Plasma,''
  Eur.\ Phys.\ J.\ C {\bf 61}, 553 (2009)
  [arXiv:0809.2869 [hep-ph]].

\bibitem{Fiol:2012sg}
  B.~Fiol, B.~Garolera and A.~Lewkowycz,
  ``Exact results for static and radiative fields of a quark in N=4 super Yang-Mills,''
  JHEP {\bf 1205}, 093 (2012)
  [arXiv:1202.5292 [hep-th]].

\bibitem{Lewkowycz:2013laa}
  A.~Lewkowycz and J.~Maldacena,
  ``Exact results for the entanglement entropy and the energy radiated by a quark,''
  JHEP {\bf 1405}, 025 (2014)
  [arXiv:1312.5682 [hep-th]].

\bibitem{Banerjee:2013rca}
  P.~Banerjee and B.~Sathiapalan,
  ``Holographic Brownian Motion in 1+1 Dimensions,''
  Nucl.\ Phys.\ B {\bf 884}, 74 (2014)
  [arXiv:1308.3352 [hep-th]].

\bibitem{Banerjee:2015fed}
  P.~Banerjee and B.~Sathiapalan,
  ``Zero Temperature Dissipation and Holography,''
  JHEP {\bf 1604}, 089 (2016)
  [arXiv:1512.06414 [hep-th]].

\bibitem{Matsuo:2006ws}
  T.~Matsuo, D.~Tomino and W.~Y.~Wen,
  ``Drag force in SYM plasma with B field from AdS/CFT,''
  JHEP {\bf 0610}, 055 (2006)
  [hep-th/0607178].

\bibitem{Kiritsis:2011ha}
  E.~Kiritsis and G.~Pavlopoulos,
  ``Heavy quarks in a magnetic field,''
  JHEP {\bf 1204}, 096 (2012)
  [arXiv:1111.0314 [hep-th]].

\bibitem{Chernicoff:2012iq}
  M.~Chernicoff, D.~Fernandez, D.~Mateos and D.~Trancanelli,
  ``Drag force in a strongly coupled anisotropic plasma,''
  JHEP {\bf 1208}, 100 (2012)
  [arXiv:1202.3696 [hep-th]].





\end{thebibliography}

\end{document}